\documentclass[dvipdfm,letter]{emulateapj} 
\usepackage{mathptmx}                      
\usepackage{hyperref}                      

\usepackage{natbib}
\usepackage{amsmath}
\usepackage{graphicx}

\shorttitle{Coronal Tomography}
\shortauthors{Charles C.\ Kankelborg}

\begin{document}

\title{Coronal Tomography}
\author{Charles C. Kankelborg}
\affil{Department of Physics, Montana State University, Bozeman, MT 59717}
\email{kankel@solar.physics.montana.edu}

\begin{abstract}
A simple, yet powerful, algorithm for computed tomography of the solar corona is
presented and demonstrated using synthetic EUV data. A minimum of three
perspectives are required. These may be obtained from \textit{STEREO}/EUVI plus
an instrument near Earth, e.g.\ \textit{TRACE} or \textit{SOHO}/EIT.
\end{abstract}


\section{Introduction}

Observations by the Extreme Ultraviolet Imager \citep[EUVI,][]{Wuelser2004} aboard
\textit{STEREO} provide the first simultaneous, stereoscopic image pairs of the solar
corona and transition region. Ideally, these data are simple projections through an
optically thin corona. However, the 3D distribution of emission is difficult to
estimate with only two projections \citep[see][and references therein]{Gary1998}. This difficulty may be explained as follows. Consider an object  $I(x,y,z)$
on domain $D$. The coordinates $x$ and $y$ need not be orthogonal, but $z$ is
orthogonal to $x$ and $y$. Given two projections $f(x,z) = \int_D I\,dy$ and
$g(y,z)=\int_D I\,dx$, a plausible reconstruction of $I$ is
\begin{equation}\label{eq:separable}
   I'(x,y,z) = \frac{f(x,z)\,g(y,z)}{T(z)},
\end{equation}
where $T(z) = \int_D I(x,y)\,dx\,dy = \int_D f\,dx = \int_D g\,dy$ is
the total emission of a plane of constant $z$. Unfortunately, this solution
fails utterly in practice. For each pair of sources in $I$,  $I'$ introduces a
pair of ``ghost'' artifacts. These are systematic errors, independent of the
noise and apparently unavoidable. Traditional regularization strategies are
not fruitful: $I'$ is positive, is as smooth as the observed images $f$ and
$g$, and is precisely the maximum entropy solution. More information is
therefore required to guide the tomographic reconstruction.

Previously described approaches to the \textit{STEREO} coronal tomography problem rely
on assumptions about the geometry of the coronal plasma distribution. The triangulation
method \citep{Gary1998,Aschwanden2005,Aschwanden2008} assumes loops with circular cross-section, and
relies on the identification of the same loops in both images. The magnetic tomography
approach \citep{WiegelmannInhester2006} also assumes loops with circular cross-section,
and incorporates magnetic field extrapolations to constrain loop geometries.  These
methods are powerful, but they require assumptions about things that one might
reasonably hope to learn from the 3D reconstruction.

I propose that EUV images taken from a third perspective---e.g.\
\textit{TRACE} or \textit{SOHO}/EIT---may provide adequate additional
constraints for coronal tomography, without any assumptions about loop
geometry or magnetic fields. I describe a simple computed tomography
algorithm, fast enough to run in real time, and demonstrate its performance
using synthetic data with three viewpoints.

\section{Algorithm} \label{sec:smart}

The Smooth Multiplicative Algebraic Reconstruction Technique (SMART) presented here has
been developed to solve a mathematically analogous problem of reconstructing spectra
for the \textit{MOSES} sounding rocket payload \citep{KankelborgThomas2001,Fox2003}.
Iterative multiplicative algebraic reconstruction techniques (MART), perhaps inspired
by the separable solution to the two-view problem (equation \ref{eq:separable}), have
been available for many years \citep{OkamotoYamaguchi1991,Verhoeven1993}.
\cite{Gary1998} used MART along with volume constraints derived from magnetic field
extrapolations to reconstruct a pair of loops from two simulated XUV images. The unique
algorithmic features of SMART are iterative smoothing and an adaptive correction
strategy. These refinements improve numerical stability and promote convergence to a
compromise between smoothness and goodness-of-fit, leading to a reduced chi-squared of
unity. 

In the $N$-view tomography problem, an object $I(x,y,z)$ on domain $D$ is
known only by $N$ projections $f_m$, taken at angles $\theta_m$:\footnote{In our coordinates, $\theta_m$ is a right-handed rotation of the object; it therefore corresponds to the eastward heliographic longitude of the observer.}
\begin{equation}
   f_m(x,z) = \int_D \mathcal{R}(\theta_m)\,I(x,y,z)\,dy.
\end{equation}
The operator $\mathcal{R}(\theta_m)$ rotates the object $I$ by angle
$\theta_m$ about the $z$ axis.\footnote{Rotation $\mathcal{R}$ could
incorporate compound angles with altitude, azimuth and roll. The extension of
SMART to the general case is straightforward.} SMART uses the projections $f_m$
to estimate $I(x,y,z)$  by the following steps: 
\begin{enumerate}
   \item \label{step:guess} Create an initial guess, $I'(x,y,z) = 1$ on $D$. 
   \item Initialize correction weights, $\gamma_m=\frac{1}{N}$.
   \item \label{step:project} $I' \leftarrow I' * K$
      (smoothing kernel $K$ defined by eq.\ \ref{eq:kernel}).
   \item Calculate projections $f'_m(x,z) = \int_D \mathcal{R}(\theta_m)\,I'\,dy$.
   \item Calculate correction factors, 
      \begin{equation} \label{eq:correctionfactor}
         C_m(x,y,z) = \mathcal{R}(-\theta_m)\,\frac{f'_m(x,z)}{f_m(x,z)}.
      \end{equation}
      Note that a nontrivial $y$-dependence is introduced through the
      rotation.
   \item \label{step:correct} Apply corrections weighted by $\gamma_m$,
      \begin{equation}\label{eq:correct}
         I' \leftarrow I' \,\prod_m C_m^{\gamma_m}.
      \end{equation}
   \item Calculate reduced chi-squared for each projection, 
      \[
         \chi_{R,m}^2 = \frac{1}{N_x N_y N_z}\sum_{xyz} \frac{(I-I')^2}{I'}.
      \]
   \item \label{step:feedback} Adjust correction weights, $\gamma_m$,
      using a control law designed to drive $\chi_{R,m}^2 \rightarrow 1$.
   \item Repeat steps \ref{step:project}-\ref{step:feedback}
      until converged.
\end{enumerate}

The $I'$ array is oversized so that rotations will not move any of the emission
outside the volume. The subphotospheric portion of the volume is zeroed in step
\ref{step:guess}, and will remain zero since the corrections are multiplicative
(eq. [\ref{eq:correct}]).

Since the correction factors (eq.\ [\ref{eq:correctionfactor}]) are ratios of
positive numbers, the reconstruction is always positive. The strength of the
$m$th applied correction is governed by $\gamma_m$. Step \ref{step:feedback}
sets up feedback to establish a dynamic equilibrium between smoothing and
correction, so that $\chi_{Rm}^2$ tends toward unity. Our control law, which has two adjustable parameters $(a,b)$, modifies $\gamma_m$ for iteration $n+1$ using a linear combination of the previous and current values of $\chi_{Rm}^2$:
\begin{align} \label{eq:control}
   \gamma_m^{n+1} &= \gamma_m^n + a\,X_m^n + b\,\Delta X_m^n,\\ 
   X_m^n &\equiv \log (\chi_{Rm,n}^2),
   \quad \Delta X_m^n \equiv X_m^n - X_m^{n-1}.
\end{align}

The algorithm is implemented in IDL, with rotations $\mathcal{R}(\theta_m)$ perfomed
via cubic  convolutional interpolation. It is possible that spurious negative voxels
could be introduced during the rotation, but negative values are eliminated from our
projections by thresholding: $f_m \ge 0$.

The normalized smoothing kernel, $K(x,y,z)$, is defined on the discrete space
of voxels as follows:
\begin{equation} \label{eq:kernel}
   K_{ijk} = \frac{\delta_{ij}\,\delta_{jk} 
      + s\,c^{(i+j+k)}}{1+s(1+4c+12c^2+8c^3)},  
      \quad c \equiv \frac{2}{5}.
\end{equation}
This form of $K$ is not crucial, but it is designed to be very nearly
isotropic. The adjustable smoothing parameter, $s \in (0,1]$, affects the
rate of convergence but has no discernible effect on the result.

\section{Synthetic Data} \label{sec:data}

\begin{figure}[htb]
   \begin{center}
      \includegraphics[width=\columnwidth]{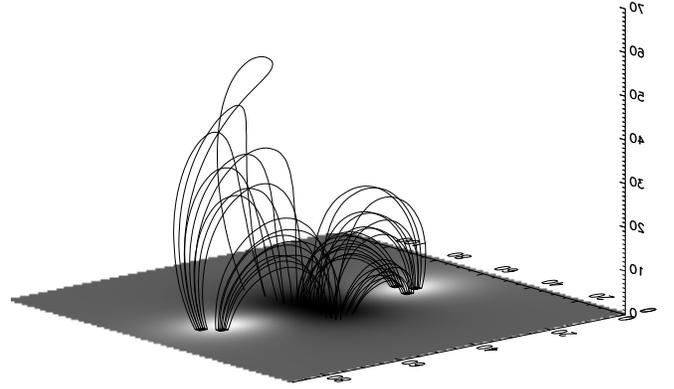}
   \end{center}
   \caption{Vertical magnetic field (shaded plane) and field lines within the emission
   from the model corona. five ``Heated'' patches  within the positive photospheric
   magnetic poles are colored in black. Note the sharp deflection of the upppermost
   field line from the coronal null.}
   \label{fig:magnetogram}
\end{figure}

\begin{figure}[htb]
   \begin{center}
      \includegraphics[width=\columnwidth]{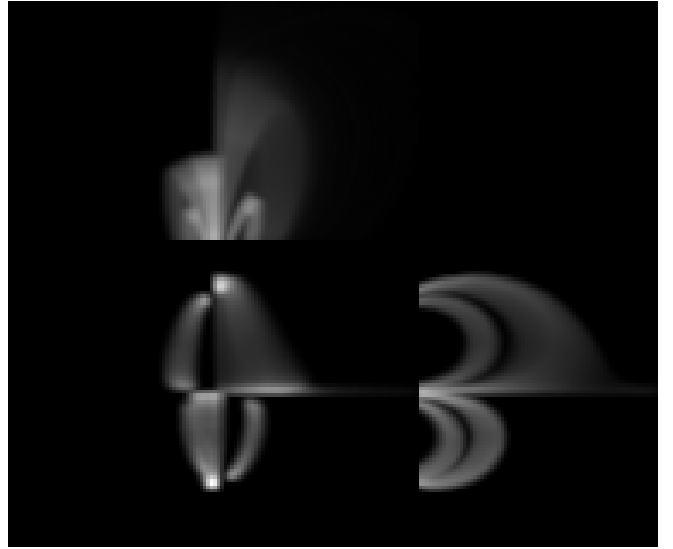}
   \end{center}
   \caption{The model coronal volume viewed along each of the three axes of
   figure \ref{fig:magnetogram}. Intensities are square-root
   scaled.}\label{fig:volume}
\end{figure}

A volume of synthetic coronal emission was prepared as a test target for the SMART
algorithm. The test target is more complex than any that have proven tractable for
previous approaches to tomographic analysis for STEREO. It is not an attempt at
detailed atmospheric modeling, but resembles a small active region. I began with a
potential field defined by four sub-photospheric magnetic charges. The resulting
line-of-sight photospheric magnetic field is shown in figure \ref{fig:magnetogram}. The
field lines in the figure emerge from five small, square ``heated'' patches in the
postive polarities on the photospheric plane. A large  number of field linese were
traced from random points in the volume to both footpoints (or to the boundary).
Emission was added to each voxel crossed by the field line, in proportion to the
arbitrary heating value at its positive photospheric footpoint. Figure \ref{fig:volume}
shows the volume model projected along each of the three axes of figure
\ref{fig:magnetogram}. The coronal flux tubes have complicated geometries, including a
broad, vertical fan of emission above the postive pole that is associated with a
coronal magnetic null at a height of $\sim 53$ grid points. All images of EUV emission
in this letter are square-root scaled to bring out faint features. 

\begin{figure*}[htb]
   \begin{center}
      \includegraphics[width=\textwidth]{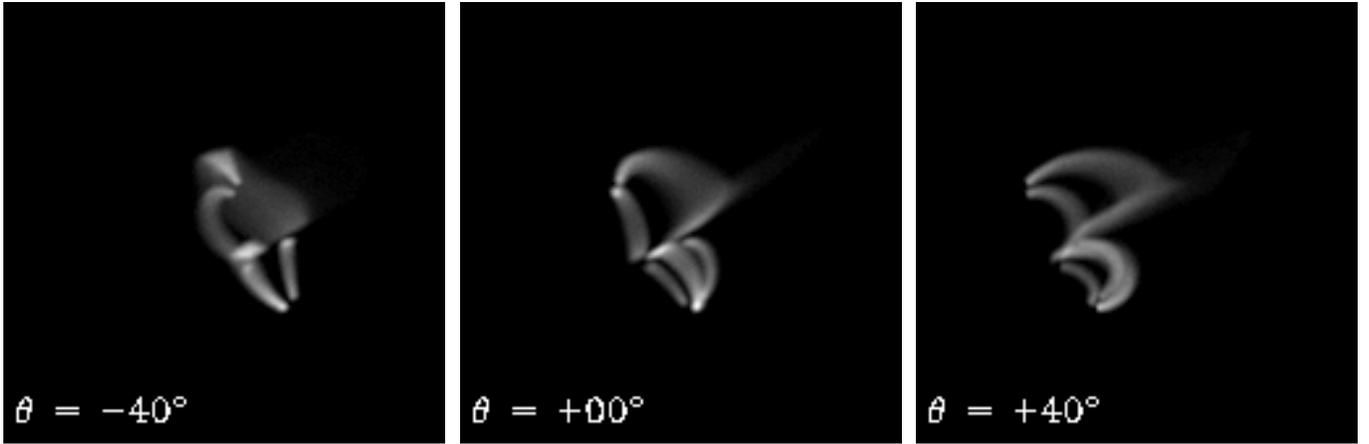}
   \end{center}
   \caption{Synthetic data taken from three virtual spacecraft. Intensities are
   square-root scaled.}\label{fig:projections}
\end{figure*}

SMART was tested on numerous synthetic observations of the model coronal volume,  each
time placing the model active region at a different random northern heliographic
latitude over $[0^{\circ},30^{\circ}]$, a random heliographic longitude over
$[-40^{\circ}, 40^{\circ}]$, and a random tilt over the interval $[-180^{\circ},
180^{\circ}]$. For simplicity, in our coordinate system the solar equator coincides
with the equatorial plane. Observations were projected for three distant virtual
instruments in the ecliptic, observing from heliographic longitudes $-40^{\circ},
0^{\circ}$, and $40^{\circ}$.  The twin \textit{STEREO} spacecraft will reach similar
separation angles in October, 2008.  The images were normalized so that the brightest
pixel among the three images had 3000 counts. The $\theta =
(-40^{\circ},0^{\circ},40^{\circ})$ projections shown in figure \ref{fig:projections}
correspond to an example observation with the region placed at latitude
$16.51^{\circ}$\,N, longitude $14.65^{\circ}$\,W, and tilt $29.63^{\circ}$
counter-clockwise. Poisson noise was applied to the images prior to passing them to the
SMART algorithm for inversion.  Intensities are square-root scaled to show the noise
more clearly. The mean value of the nonzero pixels is 216 counts.

\section{Inversion Results} \label{sec:results}

The synthetic data were inverted using 15 SMART iterations. This was typically
sufficient to converge to $\chi^2_{Rm}=1 \pm 0.01$. The rate of convergence is affected
by the smoothing parameter and by the two adjustable parameters in the control law. The
examples shown in this Letter used $s=0.5$, $a=0.05$, $b=0.16$. Within broad limits,
the results are insensitive to these parameters. For example, if the smoothing is
removed altogether, numerical instabilities arise; if it is made too strong, then it
will not be possible to reach $\chi^2_{Rm}=1$. For our computational volume of $169^3$
voxels, an inversion runs in a few minutes on a laptop computer.


\begin{figure}[!p]
   \begin{center}
      \includegraphics[width=\columnwidth]{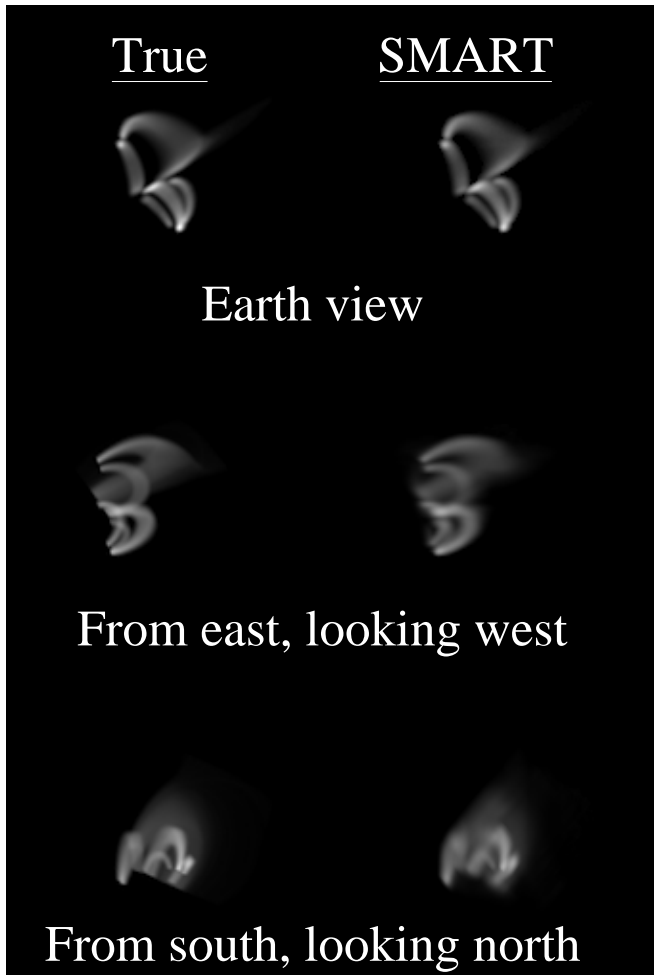}
   \end{center}
   \caption{Comparison of true and reconstructed coronal volumes from three
      orthogonal points of view. Intensities are square-root scaled.}
   \label{fig:threeview1}
\end{figure}
\begin{figure}[!p]
   \begin{center}
      \includegraphics[width=\columnwidth]{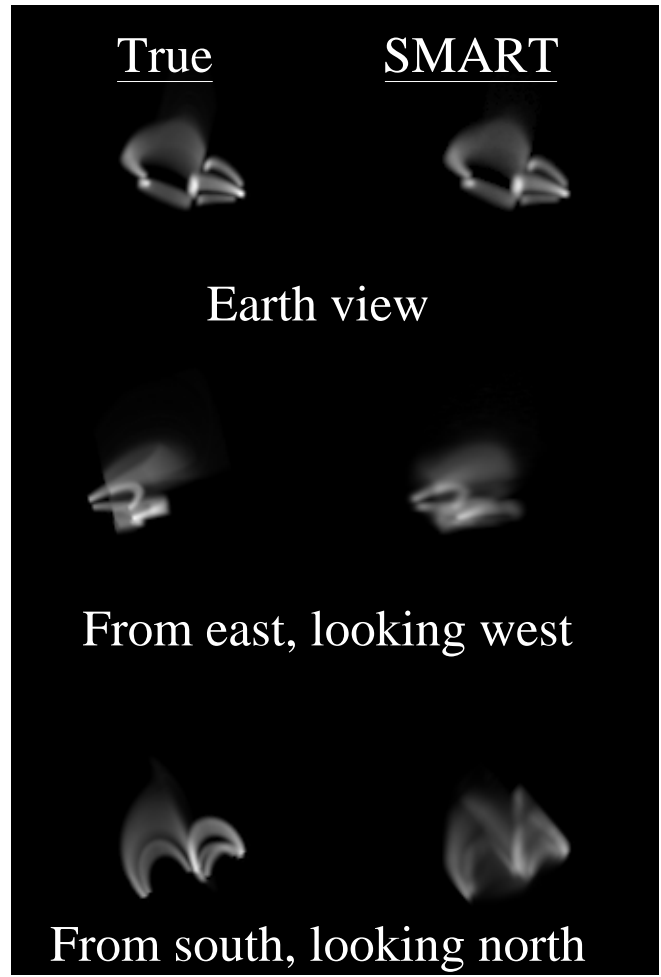}
   \end{center}
   \caption{Same as figure \ref{fig:threeview1}, but with loops
   nearly parallel to the ecliptic.}
   \label{fig:threeview2}
\end{figure}

The simulated data in figure \ref{fig:projections} gives rise to an accurate
reconstruction. 

Figure \ref{fig:threeview1} shows reconstruction results for the data in figure
\ref{fig:projections}, compared to noise-free visualizations of the coronal volume
model. The Earth viewpoint at the top of figure \ref{fig:threeview1} corresponds to the
middle panel of figure \ref{fig:projections}. Comparing these two figures shows that
the photon noise has been largely suppressed in the SMART reconstruction. The success
at recovering 3D geometry is best illustrated by the east-west and south-north
projections, which are $50^{\circ}$ and $90^{\circ}$, respectively, from the nearest
available observing angles in the synthetic data.  All of the true features have been
recovered. Square-root scaling helps to bring out the artifacts, which are few and
faint. There is slight blurring, and minimal ghosting. These results are typical of hundreds of realizations tried so far. 

A second example, with loops nearly parallel to the ecliptic plane, is given in figure
\ref{fig:threeview2}. The horizontal loop orientation is very challenging because only
the ends of horizontal features provide any depth cues.  Altough views from within the
ecliptic plane are reproduced well, the example shows relatively poor reconstruction of
an out-of-ecliptic view (lower panel). Animated versions of figures
\ref{fig:threeview1} and \ref{fig:threeview2} are provided in the electronic version of
the Journal.

\section{Discussion and Conclusions} \label{sec:discussion}

Coronal tomography is possible with as few as three viewpoints, making no prior
assumptions about coronal morphology or magnetic fields.  The test cases demonstrate
recovery of complex geometry without reference to magnetic field extrapolations or
assumptions about loop geometry. Loops that run in an east-west direction, however,
provide insufficient depth cues for three instruments confined to the  ecliptic.

The SMART algorithm described here provides a noise-insensitive tomographic
reconstruction by finding an optimal balance between goodness of fit and local
smoothness.  \textit{STEREO} will obtain data at large separation angles in Fall 2008.
Observations from \textit{SOHO} and/or \textit{TRACE} at that time should allow the
best opportunity for application of SMART to coronal tomography.

\section*{Acknowledgments}

I thank Dana Longcope for many helpful suggestions made during the preparation
of this Letter. This work was supported by NASA grant NNX07AG6G.

\clearpage

\bibliographystyle{apj}
\bibliography{apj-jour,CCKletter}

\end{document}